\documentclass{PoS}
\usepackage{amsmath,amssymb}
\usepackage{graphicx}
\usepackage{textcomp}
\usepackage{gensymb}
\usepackage{subfigure}
\usepackage{float}
\usepackage{comment}
\usepackage[numbers]{natbib}
	
\title{A Novel Analytical Model of the Magnetic Field Configuration in the Galactic Center Explaining the Diffuse Gamma-Ray Emission}

\ShortTitle{Diffuse emission from the Galactic Center}

\author{\speaker{Mehmet Guenduez}$^a$, Julia Becker Tjus$^a$, Katia Ferri\`ere$^c$, Ralf-J\"urgen Dettmar$^b$,  Dominik J. Bomans$^b$\\
	E-mail: \email{Mehmet Guenduez}\\
	$^a$Ruhr-University Bochum, Faculty of Physics and Astronomy, RAPP Center, TP IV, 44780 Bochum, Germany\\
	$^b$Ruhr-University Bochum, Faculty of Physics and Astronomy, Astronomical Institute (AIRUB), 44780 Bochum, Germany\\
	$^c$IRAP, Université de Toulouse, CNRS, 9 avenue du Colonel Roche, BP 44346, 31028 Toulouse Cedex 4, France}

\abstract{{Reliable identification of the origin of the high-energy non-thermal emission from the Galactic Center (GC) is not achievable without adequate consideration of the ambient conditions such as the magnetic field configuration or gas distribution.}
	{In a first step, we present a model that can explain the diffuse gamma-ray emission as measured by H.E.S.S. for small longitudes in the Galactic Center region but comes to grief with higher longitudes. The model is given via the solution of a transport equation that allows for a radial dependency of the mass distribution. In order to move from this semi-analytical approximation toward a full understanding of the PeVatron signature, we present a new 3D analytical model of the gas distribution in the Central Molecular Zone (CMZ). Furthermore, we derive for the first time a 3D model of the magnetic field configuration and strength in the CMZ, which is analytical and divergence-free. The model is built via a combination of a model for the diffuse inter-cloud medium, local molecular clouds and non-thermal filaments at which local information are based on investigations from previous works and the molecular gas density.}
	{It can be shown that without an efficient longitudinal CR entrapment, a single source at the center does not facilely suffice the diffuse gamma-ray detection. Further, we show that the new magnetic field model \textit{GBFD19} is compatible with recent polarization data and has a significant impact on the longitudinal profiles of CR propagation.}}

\FullConference{36th International Cosmic Ray Conference -ICRC2019-\\
	July 24th - August 1st, 2019\\
	Madison, WI, U.S.A.}

\begin{document}
	\section{Introduction}
	\noindent A diffuse $\gamma$-ray flux through the Central Molecular Zone (CMZ) has been measured by H.E.S.S. \cite{AbramowskiNature} from several GeV to tens TeV which suggests a CR spectrum extending up to the knee. Gaggero et al.(2017) \cite{Gaggero2017} presented Fermi-LAT PASS8  measurements in the CMZ at energies from $\sim$ 5 GeV to 0.26 TeV. As high-energy non-thermal emission implicates the presence of even more energetic CRs, the central vicinity of this region is most likely a cosmic ray (CR) cradle.
	However, CR electrons are not expected to cause the diffuse $\gamma$-ray emission from the inner 200 pc as they are susceptible to radiative losses and the propagation timescale is larger than the loss timescale inside the CMZ \cite{AbramowskiNature}. Still, even as of today, the real source of the CRs from the GC is not clear, and a satisfying answer can only be achieved if the ambient conditions are considered adequately.
	Therefore, the first step of this work concerns the reproduction of the high-energy (5 GeV-40 TeV) non-thermal emission regarding the energy and radial distribution via the solution of the transport equation using a semi-analytical method. In contrast, to previous works, we introduce a radial dependency into the continuous momentum loss and compare our results with the direct measurement of the $\gamma$-ray luminosity presented in \cite{AbramowskiNature} and \cite{Gaggero2017}.\\
	CR propagation is highly influenced by ambient conditions, especially by the local magnetic field and medium. For both an accurate 3D model is missing in the CMZ. Global models such as \cite{Farrar2012} always leave out the GC, and global mass distribution models do not resolve local high dense molecular clouds.
	Therefore, in the main part of this work, in order to make realistic environment more accessible for the GC investigators, we model both a 3D gas and magnetic field distribution and present the impact on CR propagation.
	\subsection*{Diffuse emission from the Galactic Center}
	\label{DEFGC}
	
	\section{Model }
	\label{Model}
\begin{figure}[H]
		\vspace*{-0.8cm}
		\centering
		\subfigure{\includegraphics[width=0.5\linewidth]{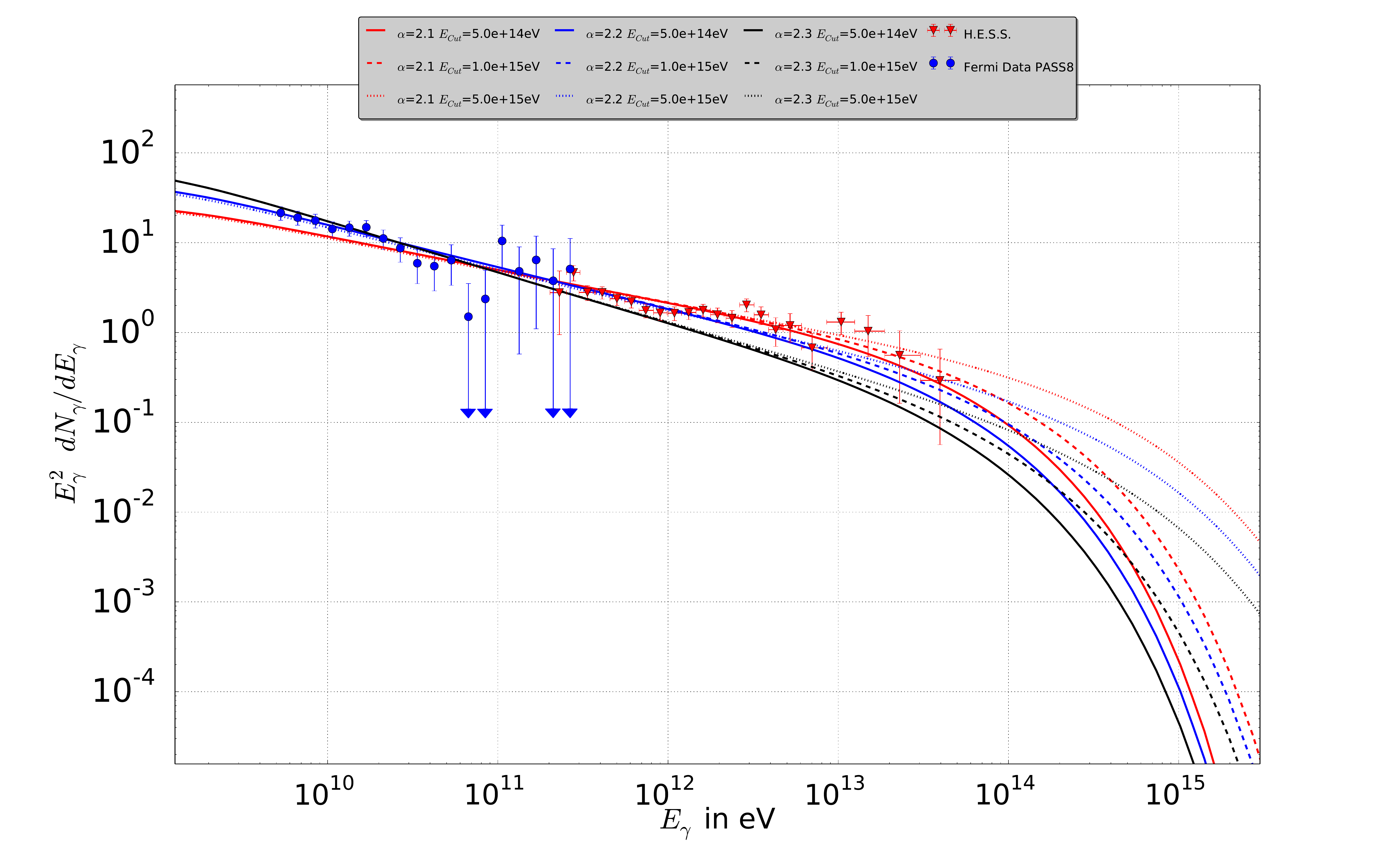}}
		\subfigure{\includegraphics[width=0.49\linewidth]{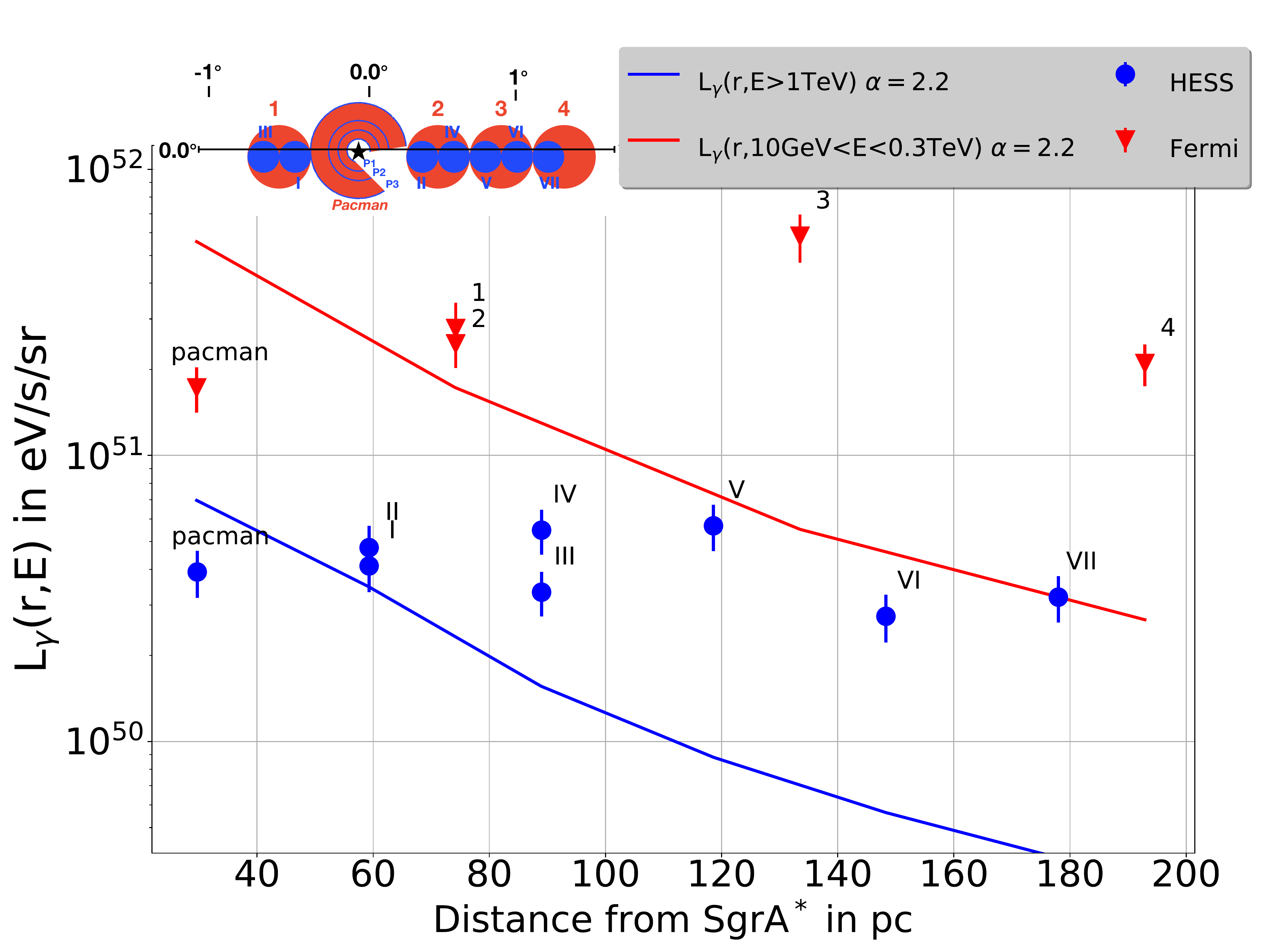}}
		\caption{Left: $\gamma$-ray energy spectrum for different source spectral indices and different cut-off energies. Right: $\gamma$-ray luminosity for $\alpha$=2.2 and $E_{\rm cut}=1$ PeV. Blue spheres are observed by H.E.S.S. and re spheres corresponds to Fermi measurements.}
		\label{GammaVariated}
	\end{figure}

\begin{figure}[H]
		\centering
		\subfigure{\includegraphics[width=0.45\linewidth]{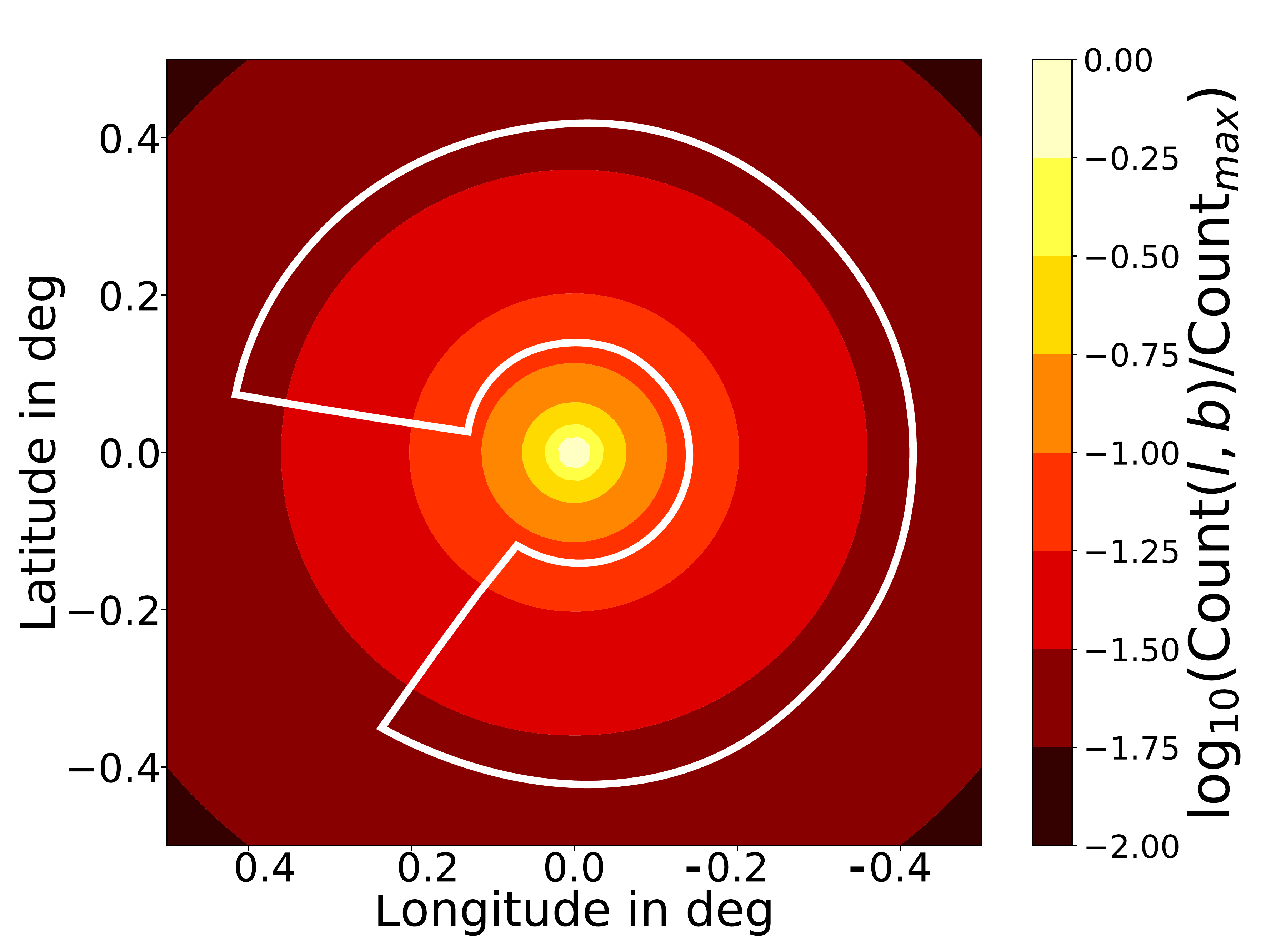}}
		\subfigure{\includegraphics[width=0.45\linewidth]{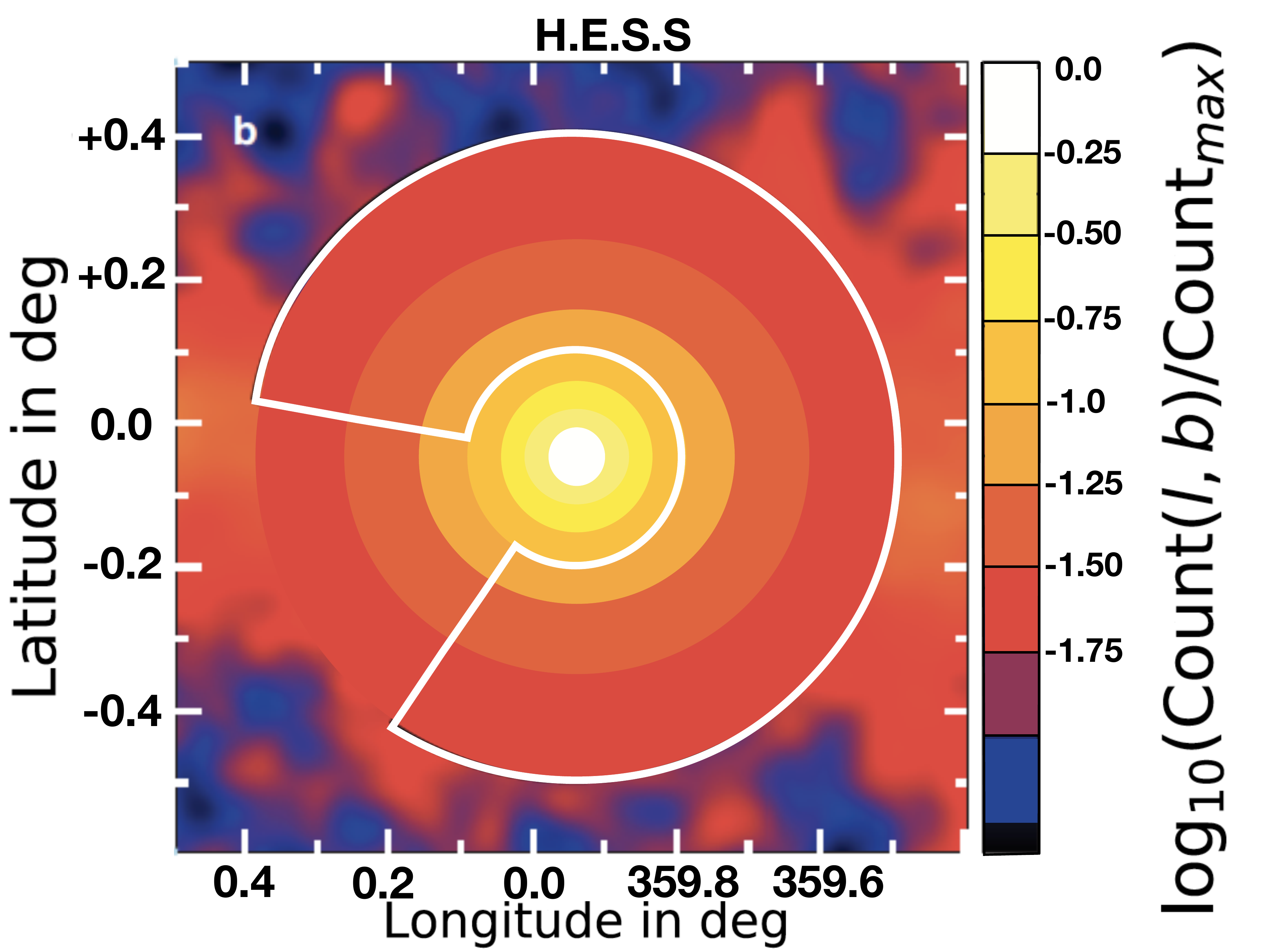}}
		\caption[]{Ratio count map from this work (left) and measured by H.E.S.S. (right)}
		\label{Count3D}
	\end{figure}

	\noindent
	We will focus our attention on proton as primary CR candidate and the radial distribution of the ambient medium, which is included in the continuous loss due to hadronic pion production \cite{SchlickiKrakau2015}. We approximate the related momentum loss rate $dE/dt=B(r,\gamma)\simeq\Lambda_{pp}\cdot N_t(r)\cdot\gamma^{1+\mu}=b(r)\cdot\gamma^{1+\mu}$ with $\mu=0.08$ and include a radial dependency due to the target density $N_t\longrightarrow N_t(r)$.
	We consider the necessity of  $\simeq1/r^2$ distribution of the CMZ to survive the tidal stresses in the steep GC potential \cite{Density1/r1.8}. We assume that at large scales, local clumps are negligible. For the first approximation, we take a density distribution of $N_t(r)=n_0\cdot (r_c/r)^2$ into account. Considering the total mass of the CMZ \cite{MassGalacticCenter}, the gas density yields $N_t(r)=n_0\cdot ({r_c}/{r})^2 \text{ with } n_0=10^4 \text{cm}^{-3}, \ r_c=2.5\cdot10^{19} \text{ cm}$.
	In the following, $b_0$ will denote $b_0=\Lambda_{pp}\cdot n_0$.
	We further assume a static and spherically symmetric description and isotropic injection of the CRs through the CMZ. This assumption is for the first approximation reasonable as until $R\approx$ 200 pc the spiral arms are not characterized. Moreover, as the advection speed is not competitive with diffusion following to \cite{AdvVern}, this process is neglected in our calculations. These settings lead to the following transport equation\\
	\begin{equation}
	-D_0\,\gamma^{\nu}\frac{2}{r}\frac{\partial}{\partial r} n(r,\gamma)-D_0\,\gamma^{\nu}\frac{\partial^2}{\partial r^2} n(r,\gamma)-\frac{\partial}{\partial\gamma} b(r)\,\gamma^{1+\mu}\,n(r,\gamma)= Q(r,\gamma).
	\label{eqq1}
	\end{equation}
	Equ. (\ref{eqq1}) describes the CR proton transport as a function of the Lorentz factor $\gamma$.
	Here, we use the Kolmogorov's turbulence spectrum with $\nu =1/3$ and the average galactic diffusion coefficient $D_0=6\cdot10^{28}$ cm$^2$/s.
	Equ. (\ref{eqq1}) is solved by using the Greens method, and further by applying the \text{Duhamel's principle} in order to separate the transport equation in a momentum and radial part.
	In addition, Laplace transformations in combination with the variation of constant method, are performed individually in the momentum, as well as in the radial part.
	In doing so, the solution for an arbitrary source distribution is given by
	\vspace*{-0.5cm}
	\begin{equation}
	\begin{split}
	n(r,\gamma)=&\gamma^{-1-\mu} \int\int 
	\frac{r_0^{\frac{3}{2}} H[r-r_0]\cdot H[\gamma_0-\gamma]}{\sqrt{4\pi \cdot D_0\cdot b_0 \cdot r_c^2}}\cdot \frac{ \gamma_0^{\mu+2\nu}}{\sqrt{\frac{\gamma^{\nu}-\gamma_0^{\nu}}{\nu}}}\cdot\exp\left( -\frac{D_0}{4\nu\cdot b_0\cdot r_c^2}(\gamma^{\nu}-\gamma_0^{\nu})\right) \cdot \\
	&\left[\exp\left( -\frac{\log(\sqrt{r_0})^2\nu\cdot b_0\cdot r_c^2}{D_0\cdot (\gamma^{\nu}-\gamma_0^{\nu})}\right) -\exp\left( -\frac{\log(\frac{r}{\sqrt{r_0}})^2\nu\cdot b_0\cdot r_c^2}{D_0\cdot (\gamma^{\nu}-\gamma_0^{\nu})}\right) \right] 
	\cdot Q(r_0,\gamma_0)\rm  d r_0 \rm d\gamma_0
	\end{split}
	\label{nn-sol-general}
	\end{equation}
	\paragraph{Calculation of the gamma-ray emission}
	The $\gamma$-ray flux from hadronic pion production, which is supposed to be the primary source of the diffuse $\gamma$-ray emission from the GC can be calculated as follows \cite{Guenduez18}:
	\vspace*{-0.5cm}
	\begin{equation}
	\begin{split}
	&\Phi_{p,\pi^0,\gamma}^{had}(E_{\gamma})=\frac{c}{ d^2}\int_{r_{min}}^{r_{max}}\int_{E_{\gamma}}^{\infty} \sigma_{pp, inel}(E_p) \cdot N_t(r)\cdot n_p(r,\frac{E_p}{m_p})F_{\gamma}(\frac{E_{\gamma}}{E_{\pi}},E_p) \frac{dE_p}{E_p}\cdot r^2\rm d r \\
	&L_{\gamma}(E_{min},E_{max})=\int_{E_{min}}^{E_{max}}\Phi_{p,\pi^0,\gamma}^{had}(E_{\gamma}) E_{\gamma} \cdot A\cdot  {\rm d}E_{\gamma}\, .
	\end{split}
	\label{Luminosity1}
	\end{equation}
	Relevant energy spectra and cross sections from inelastic proton-proton interaction are taken from \cite{Kelner}. Furthermore, we consider a CR source extended from the Schwarzschild radius $R_s$ to the maximum range of the accretion zone $r_{\rm a}=3\cdot10^3\cdot R_s$ \cite{AccretionSgrA}, which obeys a power law distribution in the energy and includes an exponential energy cut-off: $Q(r,\gamma)=q_0\cdot \gamma^{-\alpha}\cdot \exp(-\gamma/\gamma_{\rm cut})\cdot H[R_{a}-r]H[r-R_{s}]
	$.
	We keep the spectral index $\alpha$,  cut-off Lorentz factor $\gamma_{\rm cut}$, and normalization factor $q_0$ as free parameters.
	Moreover, we are interested in the explanation of the radial profile of the $\gamma$-ray luminosity measured by  H.E.S.S. and Fermi. For this purpose, the second line of Equ. (\ref{Luminosity1}) calculates the expected luminosity from the $\gamma$-ray flux with $A$ as the surface of the region of interest.
	\vspace*{-0.5cm}
	\paragraph{Comparison of data with results}
	\noindent First of all, we determine all free parameters by fitting them to the differential flux from the inner $\sim67$ pc as declared as the \textit{Pacman} region in \cite{AbramowskiNature}. Figure \ref{GammaVariated} displays the resulting $\gamma$-ray fluxes for different spectral indices and cut-off energies.
	The best fit procedure delivers a spectral index of $\alpha=2.2$ and does not distinctly prefer specific cut-off energy between 0.5-5 PeV due to the large measurement errors. However, as galactic accelerator supposed to reach energies up to the known \textit{knee}, we rely on 1 PeV.
	The H.E.S.S. data of the spatial $\gamma$-ray distribution are reproduced at high precision in the \textit{Pacman} region given in Figure \ref{Count3D}. The agreement might succeed due to the spherical structure in the Pacman region as assumed in our model. 
	The predicted luminosity progression at H.E.S.S. energies is more inconsistent with the data, the farther we move away from the point of origin. This effect becomes stronger at Fermi energies and might indicate additional unconsidered high-energy sources at higher longitudes or a complex magnetic field configuration as higher energetic CRs are less deflected.
	\vspace*{-0.5cm}
	In order to move away from this semi-analytical approximation toward a full understanding of the PeVatron signature, we further present a new 3D model of both gas and magnetic field distribution in the CMZ.
	\ShortTitle{Gas distribution in the Galactic Center}
	\section{Gas distribution}
	\noindent Concerning the CR propagation, the gas distribution is for two purposes essential as on the one hand, due to inelastic p-p interaction pions are generated which decay among others into $\gamma$-rays. On the other hand, the gas density inside the MCs in the CMZ has enmeshed with the magnetic field strength a consequence of magnetic flux conservation. Wardle and K\"onigl (1990) \cite{CNRMagneticField} provide the relation for the MCs near the GC following
	$\lvert \textbf{B}_{\rm eq}\rvert=1.5\text{ mG}\cdot (\frac{n}{10^4 \text{ cm}^{-3}})^{{1}/{2}}
	\label{B-n-Relation}$.
	Although Ferri\`ere et al. (2007) \cite{MassGalacticCenter2} presented the average gas density distribution in the CMZ, this model does not include local structures. Therefore, this model will be considered in the diffuse intercloud (IC) medium (ICM) region. The total density profile of the GC region is composed out of three individual components: 1. local MCs whose position, size and density can be taken from Table 1 in \cite{GMDF2019}; 2. inner 10 pc which is described in \cite{Inner10pc}; 3. ICM whose analytical description is given in \cite{MassGalacticCenter2} and where the normalization factor is scaled down by the subtraction of the sub-structures to $n_{0,H_2}=128.4$ cm$^{-3}$ and $n_{0,H}=7.5$ cm$^{-3}$. 
	The combined gas density distribution is visualized in Figure \ref{MassDistr1}. In the following figures, the red star denotes the position of SgrA*.
	\ShortTitle{Magnetic field in the Galactic Center}
	\section{Magnetic field configuration}
	\noindent There are many models describing the Galactic magnetic field. However, all of them leave out the GC region although a large fraction of the CR flux is supposed to come from there. 
	In order to solve this lack and give a description of the magnetic field in the central 200 pc, we use existing pieces of information.
	The total magnetic field will be given via a combination of a model for: (1.) the diffuse ICM,  (2.) local non-thermal filament (NTF) and (3.) local MC regions. 
	Briefly, the background magnetic field in the ICM and the NTF regions is predominantly poloidal, whereas the horizontal field dominates in the MC regions. Further, an average value of $\overline{B}_{\rm IC}\approx$10 $\mu$G determines the field strength in the ICM \cite{FerriereMagneticField2009}. The magnetic field strength in the MC and NTF region can be taken from Table 1 in \cite{GMDF2019}. The total field in the CMZ is then obtained by a superposition of the magnetic fields which are derived in the next subsections :
	$
	\textbf{B}_{\text{tot}}=\textbf{B}_{\text{IC}}^C+\sum_{i=1}^{8}\textbf{B}_{\text{NTF,i}}^C+\sum_{i=1}^{12}\textbf{B}_{\text{MC,i}} 
	$.
	
	\paragraph{Poloidal field}
	\label{poloidal}
	\noindent An analytical divergence-free and isotropic poloidal field model  (\textit{Model C (FT14-C)}) which is taken from \cite{X-ShapeModel}, will describe the poloidal field in the IC and NTF region.
	\begin{eqnarray}
	\textbf{B}^C=\begin{pmatrix}
	B_r\\B_{\phi}\\B_z
	\end{pmatrix}=
	\begin{pmatrix}
	\frac{2\, a\, \, z}{(1+a\, z^2)^3} \nonumber\\
	0\nonumber\\
	\frac{1}{(1+a\, z^2)^2}
	\end{pmatrix}\cdot B_1\cdot e^{-r/L \cdot\frac{1}{(1+a\,z^2)}}
	\label{xshape:equ}
	\end{eqnarray}
	{IC region:}
	In the ICM region, the exponential scale length $L=158\text{ pc}/{\ln(2)}=114 \text{ pc}$ is adapted to the CMZ geometry following Ferri\`ere et al.\ (2007) \cite{MassGalacticCenter2} and considering cylindrical geometry. The parameter $a$ governing the opening of field lines away from the z-axis is fitted to the polarization map of Nishiyama et al. (2010) \cite{Nishiyama} in a region where no NTFs and MCs are located and results in $a=1/(42 \ \rm pc)^2$. The normalization factor $B_1$ is determined by calculating the average value of $\textbf{B}^C$ and equating to the observed and expected field strength $\overline{B}$, which yields  $B_1=\overline{B}_{\rm IC}/{\alpha}\approx 12\,\mu{\rm G}$.\\\\
	{NTF regions:}   \ \ In the NTF regions, $L$ and $a$ are given by the total longitudinal extend $\Delta l$ and $\Delta b$ of the NTFs following $L=\Delta l/{(2\cdot \ln(2))}$ and $1/{\sqrt{a}}={\Delta b}/{(2\cdot\sqrt{\ln(2)})}$. The value of $\alpha$ corresponding to the normalization factor  $B_1$ results in all NTFs approximately the same and yields $B_1\approx{\overline{B}_{\rm NFT}}/{0.26}\,$.
	\paragraph{Horizontal field}
	\label{horizontal}
	\noindent As the magnetic field in the dense MCs is predominantly parallel to the Galactic Plane, we will neglect the $B_z$ component and regard, therefore, the zonal plane. 
	The relation between the azimuthal component $B_{\phi}$ and radial component $B_r$  depends on the characteristics of the MCs, such as the gas density and the intrinsical rotational velocity. The relation in the circumnuclear disc cloud is given in \cite{CNRMagneticField} and is also adaptable to other MCs in the GC as all of them exhibits comparable high gas densities and rotational velocities. The resulted ratio $B_r/B_{\phi}=\eta$ of each MC is listed in Table 1 in \cite{GMDF2019} as well. In order to consider the conditions mentioned above in this section, we will derive a suitable field model by considering the Euler's $\alpha$ and $\beta$ potential. These potentials will deliver a divergence-free expression of the magnetic field following $\vec{B}=\vec{\nabla}\alpha\times\vec{\nabla}\beta=(B_r,B_{\phi},0)^T$ with $\beta=z$ in cylindrical coordinates naturally. 
	\begin{equation}
	\vec{B}=\begin{pmatrix}
	B_r\\
	B_{\phi}\\
	0
	\end{pmatrix}=\begin{pmatrix}
	\frac{1}{r} \frac{\partial}{\partial{\phi}}\alpha\\
	-\frac{\partial}{\partial r}\alpha\\
	0
	\end{pmatrix}=
	\begin{pmatrix}
	\frac{1}{r} \frac{\partial \alpha}{\partial{\psi}}\frac{\partial\psi}{\partial \phi}|_r \\
	-\frac{\partial \alpha}{\partial{\psi}}\frac{\partial\psi}{\partial r}|_{\phi} \\
	0
	\end{pmatrix}
	=
	\begin{pmatrix}
	\frac{\rho}{r}\,\xi(\psi)\cdot h(z)\\
	- \eta^{-1}\cdot\frac{\rho}{r+b}\, \xi(\psi)\cdot h(z) \\
	0
	\end{pmatrix}
	\label{bfield_azim:equ}
	\end{equation}
	Here, ${B_{r}}/{B_{\phi}}=\frac{1}{r} \cdot{dr}/{d\phi} |_{\psi,\rho}\overset{!}{=}\eta$ leads to $\psi\approx\phi\pm \eta^{-1}\,\ln\left( {r+b}/{\rho+b}\right)$ and we define $\partial \alpha/\partial \psi:= \rho\cdot \xi(\psi)\cdot h(z)$. $\psi$ and $\rho$ denote a reference angle and radius, respectively, and thus $\rho$ will set to the MC radius. $b$ is a free parameter which avoids the singularity of the $\ln$ function at $r=0$ and therefore will set to a negligible fraction of the MC radius.
	To ensures the physical reality of a magnetic field, i.e., the net magnetic flux is zero, we set $\xi(\psi)=B_1\cdot\cos({\psi})$. In doing so, the $h(z)$ is arbitrary, which we set to a Gaussian vertical distribution. The vertical scale length $H_c$ is then given by $H_c=\rho/\sqrt{2}$. Assembling all relations in Equ. (\ref{bfield_azim:equ}) would lead to a singularity due to the factor $\propto 1/r$. In the physical context, here, all field lines meet at $r=0$ due to this expression. We prevent the singularity within a small radius $r'$ which should correspond to a negligible fraction of the MC radius, by introducing a new $\phi$ component. The additional $\phi$ component connects two field lines instead of allowing them to meet each other.
	Then, the magnetic field in a MC is given by Equ. (\ref{BFieldModified})
	\vspace*{-0.2cm}
	\begin{equation*}
		\vec{B}^{r>r'}_{\text{MC}}=\beta
		\begin{pmatrix}
			\frac{R}{r} \\
			\mp\eta^{-1}\frac{R}{r+b}  \\
			0
		\end{pmatrix}
		\text{ with } v(r)=\frac{1}{\eta}\cdot\ln\left( \frac{r+b}{R+b}\right), \beta=B_1\cos\left(  \pm v(r) + m{\phi}\right) \cdot\exp\left(- \frac{z^2}{H_c^2} \right)
	\end{equation*}
	\vspace*{-0.7cm}
	\begin{equation}
	\vec{B}^{r<r'}_{\text{MC}}=\beta\cdot\frac{R}{r'}\left( \frac{3r}{r'} -\frac{2r^2}{r'^2}\right)
	\begin{pmatrix}
	1\\
	\mp \frac{r}{\eta(r+b)}\left( 1 +\frac{6(r-r')}{2\, r-3\, r'}\left(\frac{\sin(\pm v(r)+m\phi)-\sin(\pm v(r))}{\cos(\pm v(r)+m\phi)}\right)\right)  \\
	0
	\end{pmatrix}.
	\label{BFieldModified}
	\end{equation}
	\vspace*{-0.5cm}
	For more Details on the magnetic field modeling in the GC, see \cite{GMDF2019} (submitted).
	\section{Results}
\begin{figure}[H]
		\centering
		\subfigure{\includegraphics[width=0.8\linewidth]{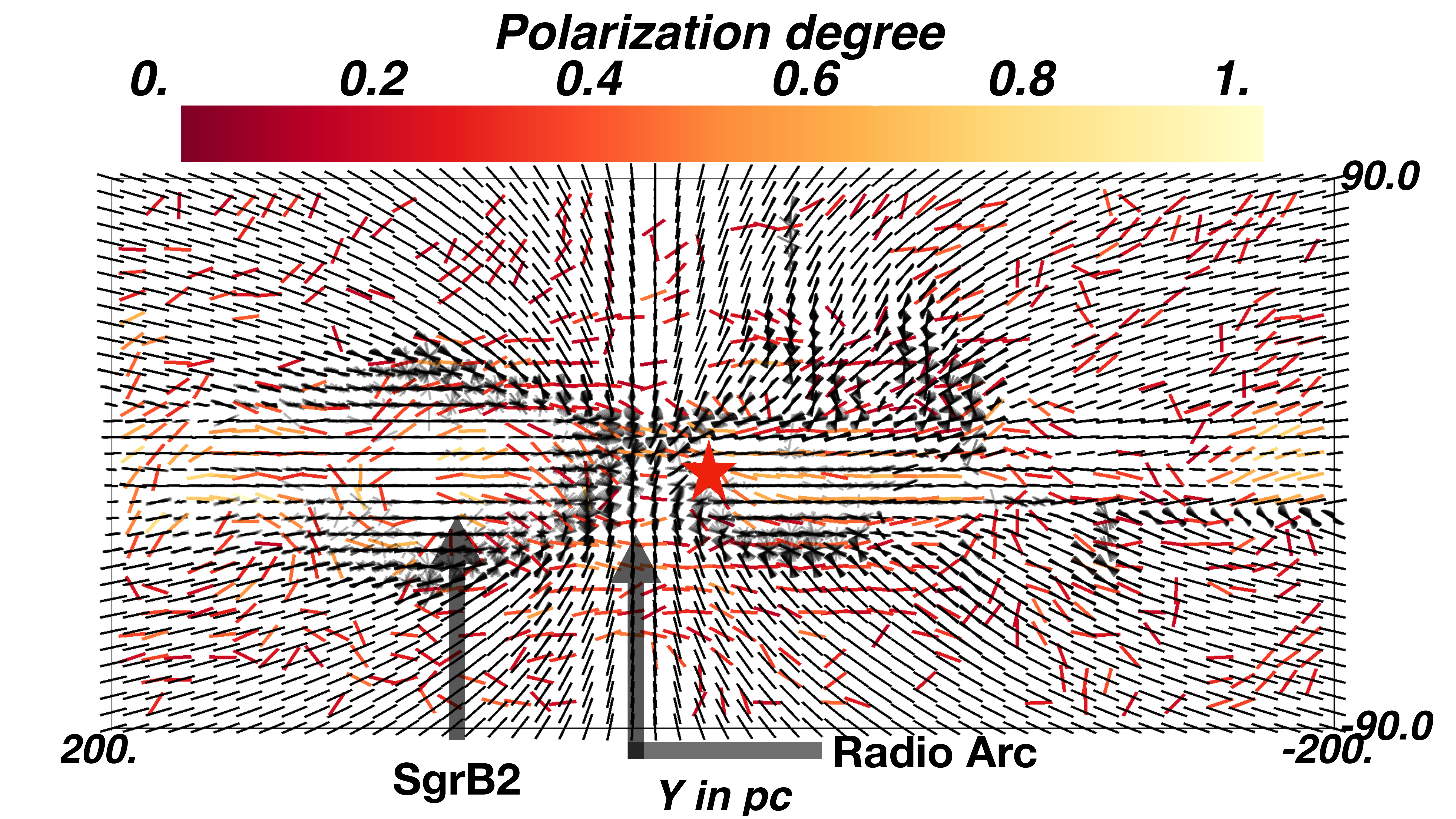}}
		\caption[]{Magnetic field configuration as derived in this work is visualized by black dashed lines and as measured by \cite{Nishiyama} by colored dashed lines. Blurred black configurations indicate polarization degree with respect to the $x$ direction.}
		\label{Pol}
	\end{figure}

	\ShortTitle{Results}
	\begin{figure}[H]
		\vspace*{-1.0cm}
		\centering
		\subfigure[Combined gas density of the CMZ region: the diffuse ICM is shown as gray contours and the contour represents a density level.]{\includegraphics[width=0.45\linewidth]{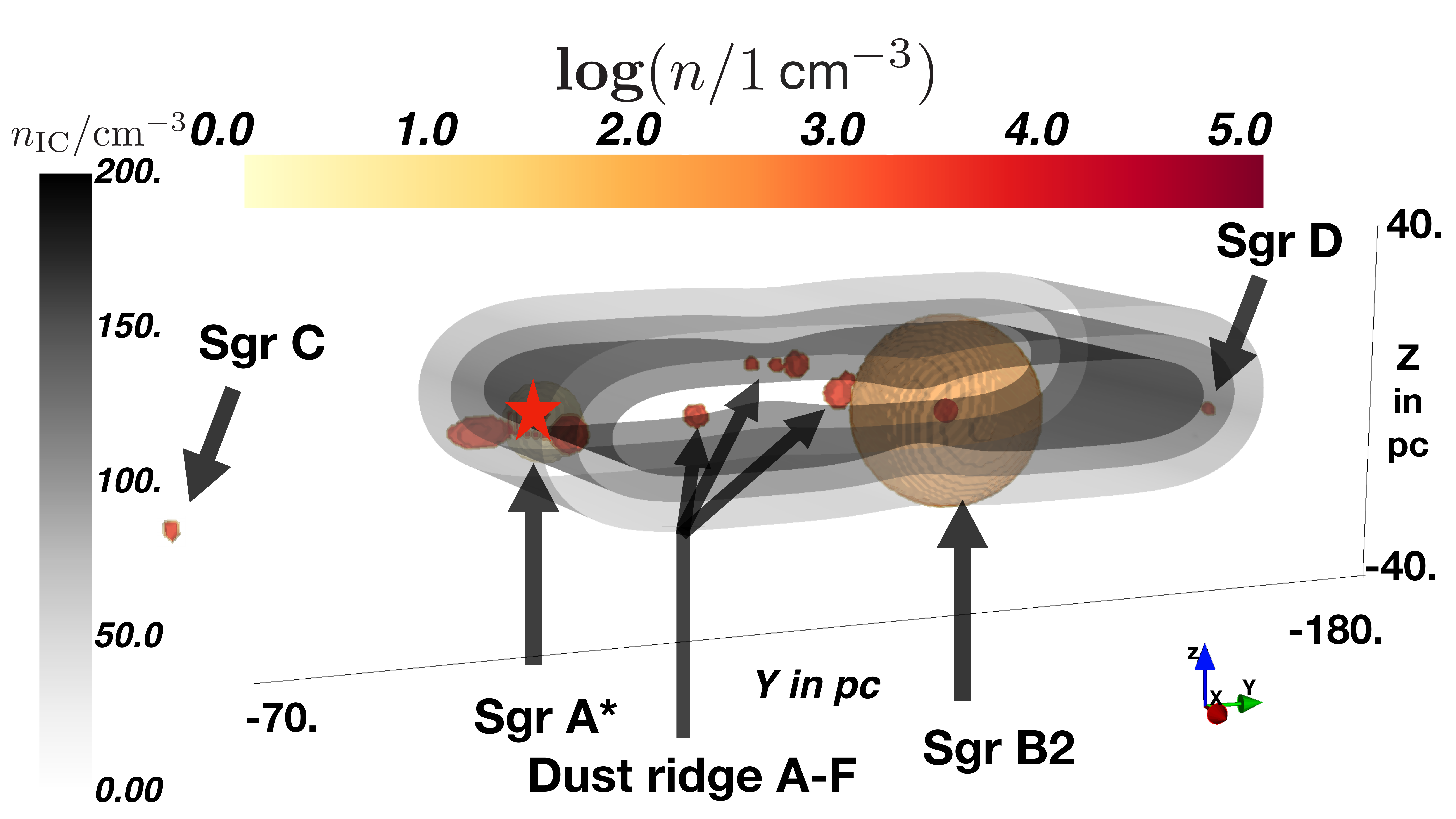}\label{MassDistr1}}
		\subfigure[The total magnetic field strength in the CMZ is given in $\mu$G. Additionally, NTFs a visualized by black cylinders.]{\includegraphics[width=0.54\textwidth]{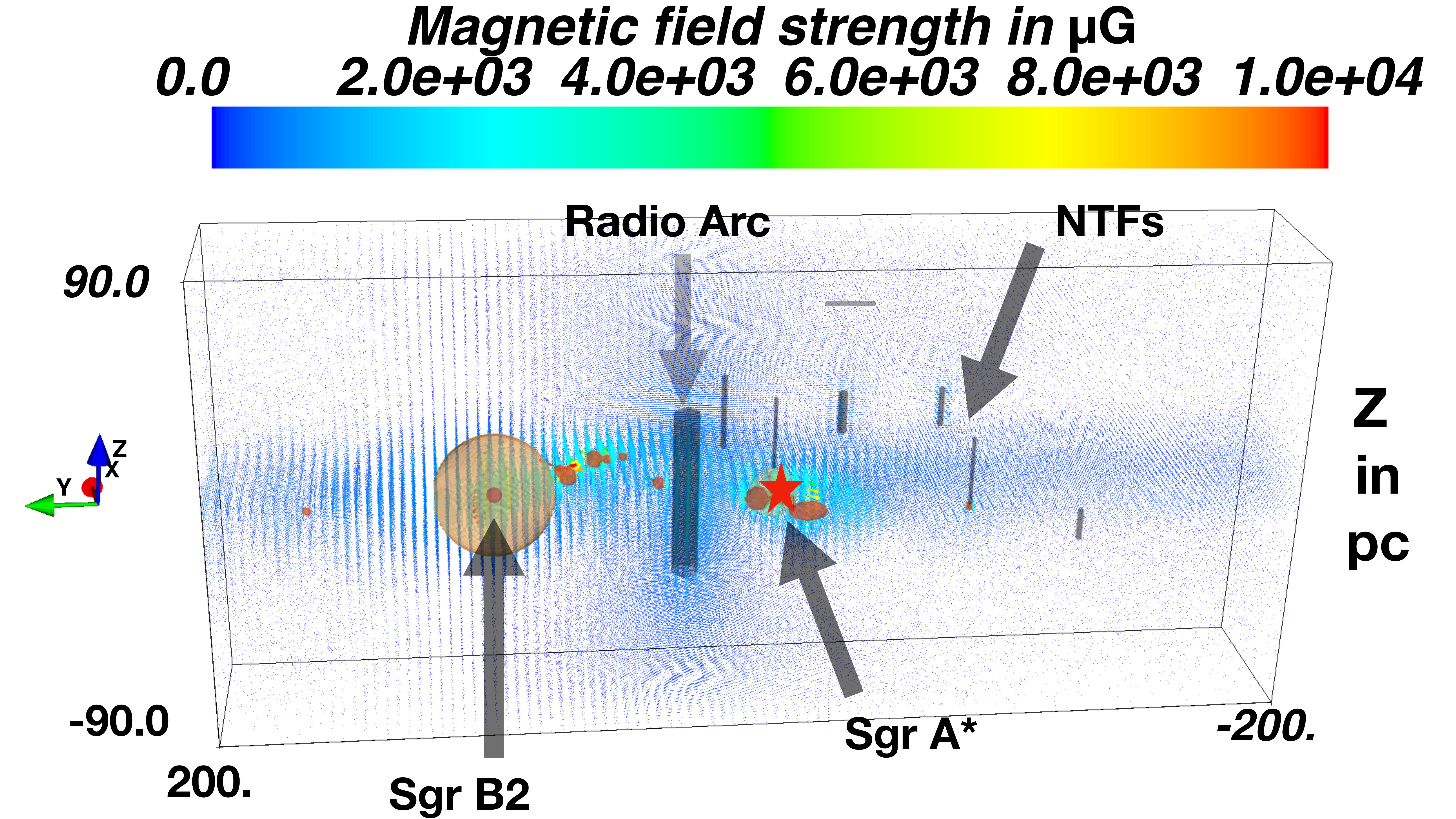}\label{FieldTotalWithObjects}}
	\end{figure}
	\vspace*{-0.5cm}
	\noindent Figure \ref{FieldTotalWithObjects} displays the 3D configuration and strength of the total magnetic field in the CMZ. The polarization map of this total field and the measurements in the CMZ by \cite{Nishiyama} are shown in Figure \ref{Pol} in order to cross-check the results. In total, the most relevant structures in the region $|l|<1.35\degree$ and $|b|<0.6\degree$ are well described by the model derived in this work. Thus, regions with good conformity, e.g., $|b|<0.3 \degree$ indicate a dominant regular magnetic field.
	\ShortTitle{References}
	\paragraph{Impact on cosmic ray propagation}
	In order to test our magnetic field model on CR propagation, we consult the CR propagation tool CRPropa 3.1 \cite{CRpropa2017}. Skipping all technical detail, we present the result of the model build in this work (\textit{GBFD19}) and the standard field used in cosmic-ray propagation presented in Jansson and Farrar (2012) \cite{Farrar2012} (abbreviated as \textit{JF12} in the following). By using the mulit-particle picture of CRPropa tool, we construct an $234\times234 \times234$ pc$^3$ environment, inject $10^5$ particles at energies 1 TeV- 1000 TeV in center and detect each particle after a discrete time step of $\Delta t_i=1$~pc/c for 234 times. 
	As a first result, at a distance  $Y=\pm100$~pc and $Z=0$~pc, the \textit{JF12} configuration results in a flux level close to zero particles, whereas the GBFD19 field reaches a level of $10^3$ protons.
	The entrapping ability of the \textit{GBFD19} could explain the longitudinal extension up to 220 pc of the diffuse $\gamma$-ray emission detected by H.E.S.S.. 
	\begin{figure}[H]
		\centering
		\hspace*{-0.cm}
		\subfigure{\includegraphics[width=0.4\linewidth]{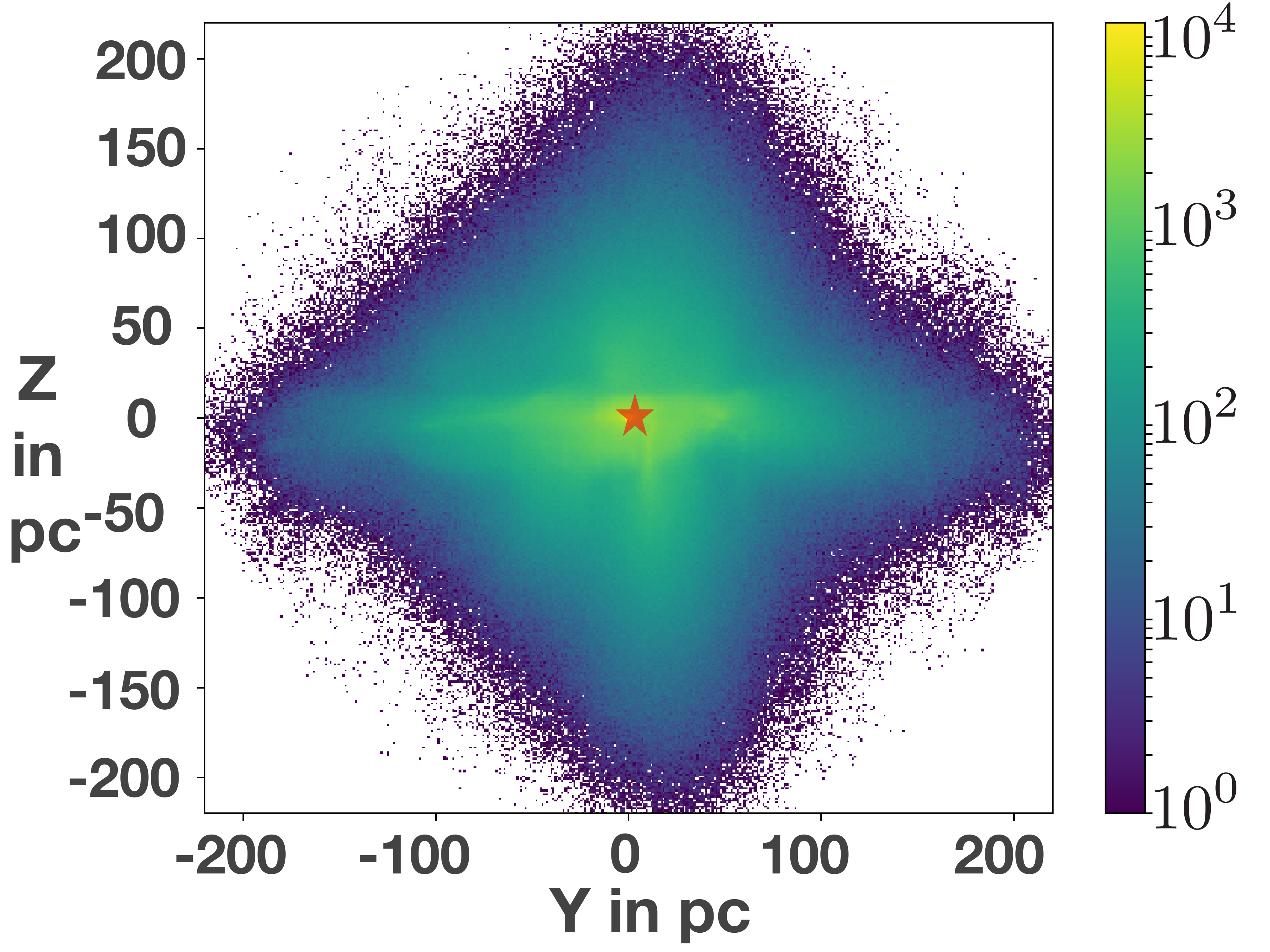}}
		\subfigure{\includegraphics[width=0.4\linewidth]{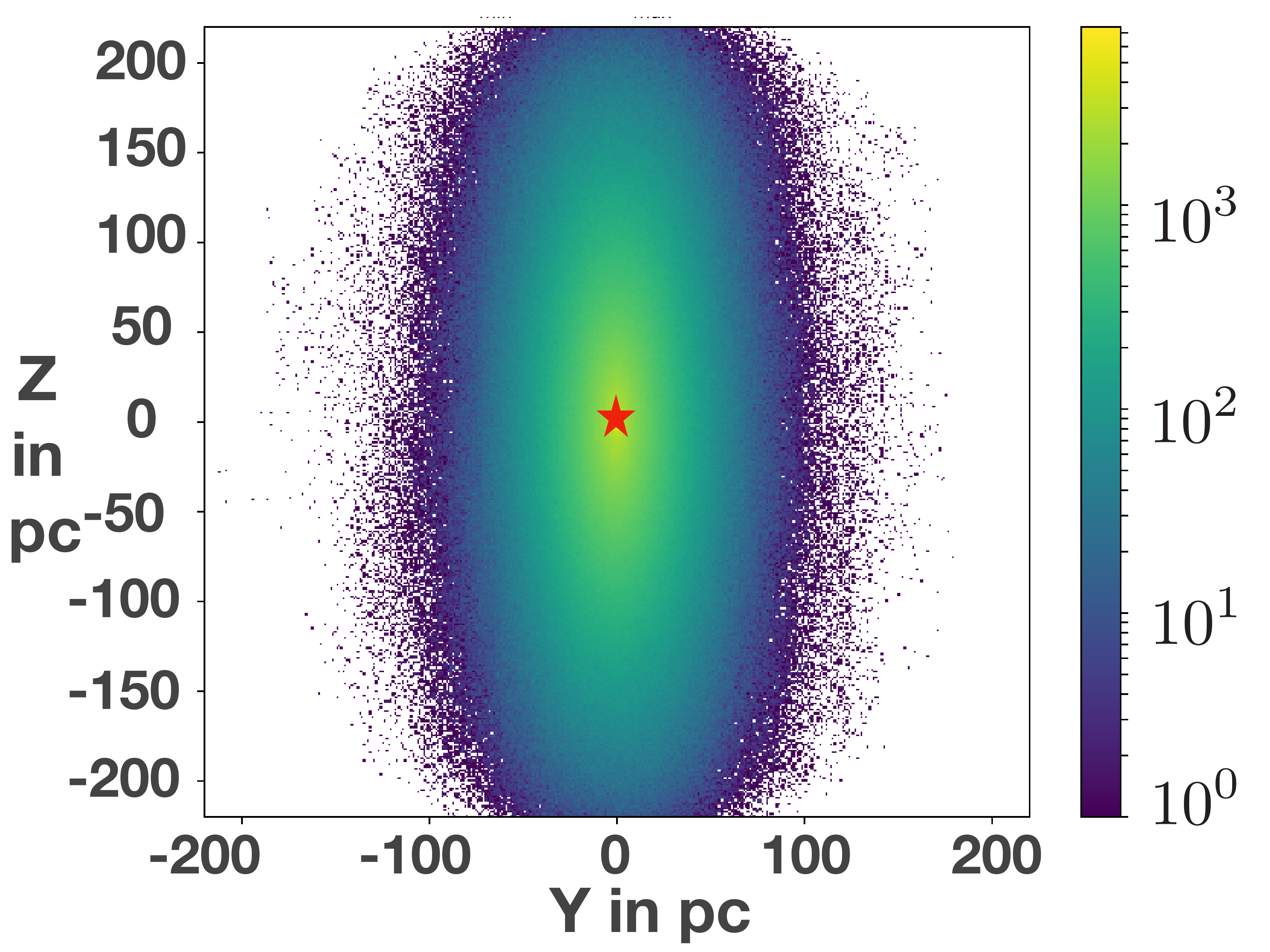}}
		\caption[]{Protons propagated in the \textit{GBFD19} (left) and in the \textit{JF12} (right) field between $1-1000$~TeV.}
		\label{Traj}
	\end{figure}
	
	\section{Summary and conclusion}
	\noindent In the first part of this work, we consider approximated ambient conditions of the GC in the transport equation (\ref{eqq1}). Hereafter, we present the solution and calculate the diffuse $\gamma$-ray emission as described in Abramowski et al. (2016) \cite{AbramowskiNature}. Furthermore, we calculate the radial distribution of the $\gamma$-ray luminosity and compare with the data presented in Gaggero et al. (2017) \cite{Gaggero2017} and Abramowski et al. (2016) \cite{AbramowskiNature}. We assert that under our assumptions, the injection of a centralized source is not able to explain the diffuse $\gamma$-ray emission at higher longitudes. Therefore, we are aimed at modeling an accurate 3D gas and magnetic field distribution. For the former modeling, we combined a 3 component model made by a diffuse ICM, local MC and by the inner 10 pc around SgrA*. The magnetic field is separated into 3 regions: 1. IC  2. NTF  3. MC region where the first 2 regions are described by a poloidal field model which is taken from \cite{X-ShapeModel}. The latter region is described by a horizontal model which is exclusively derived and describe each MC individually by considering the ambient conditions such as the gas density and intrinsic rotational velocity. The total magnetic field is then given by a superposition of the  magnetic field of each region. The polarization map result is in a strong agreement with the data measured by \cite{Nishiyama}. The influence of this field on CR propagation exhibits a significant impact. In fact, the field structure in the CMZ produce an entrapment of CR in longitudinal direction and therefore is most likely responsible for the longitudinal extension up to 220 pc of the diffuse $\gamma$-ray emission detected by H.E.S.S.. However, for a quite validation further simulation of the production $\gamma$-ray in an realistic environment has to be performed. The 3D gas and magnetic field distribution presented in this work makes this goal more accessible to the GC investigators.
	\bibliography{literaturNew}
	\bibliographystyle{unsrt}
\end{document}